\documentclass[aps,twocolumn,preprintnumbers,showpacs,showkeys,nofootinbib%
]{revtex4-1}
\pdfoutput=1

\usepackage[T1]{fontenc} 
\usepackage[utf8]{inputenc}

\usepackage{epsfig}
\usepackage{amssymb,amsmath,amsfonts,amsthm,graphicx,psfrag}
\usepackage{hyperref}

\newcommand{\beq}{\begin{eqnarray}}
\newcommand{\eeq}{\end{eqnarray}}

\newcommand{\tr}{\operatorname{Tr}}

\newcommand{\roundcustom}{\operatorname{round}}

\begin{document}
\title{Lattice study of QCD at finite chiral density: topology and confinement}

\author{N.\,Yu.~Astrakhantsev}
\email[]{nikita.astrakhantsev@phystech.edu}
\affiliation{Physik-Institut, Universität Zürich, Winterthurerstrasse 190, CH-8057 Zürich, Switzerland\\
Moscow Institute of Physics and Technology, Institutsky lane 9, Dolgoprudny, Moscow region, 141700 Russia\\
Institute for Theoretical and Experimental Physics NRC ``Kurchatov Institute'', Moscow, 117218 Russia}

\author{V.\,V.~Braguta}
\email[]{braguta@theor.jinr.ru}
\affiliation{Bogoliubov Laboratory of Theoretical Physics, Joint Institute for Nuclear Research, Dubna, 141980 Russia\\
Far Eastern Federal University, School of Biomedicine, 690950 Vladivostok, Russia}

\author{A.\,Yu.~Kotov}
\email[]{andrey.kotov@phystech.edu}
\thanks{Corresponding author}
\affiliation{Moscow Institute of Physics and Technology, Institutsky lane 9, Dolgoprudny, Moscow region, 141700\\
Institute for Theoretical and Experimental Physics NRC ``Kurchatov Institute'', Moscow, 117218 Russia\\
Bogoliubov Laboratory of Theoretical Physics, Joint Institute for Nuclear Research, Dubna, 141980 Russia}

\author{D.\,D.~Kuznedelev}
\email[]{kuznedelev.dd@phystech.edu}
\affiliation{Moscow Institute of Physics and Technology, Institutsky lane 9, Dolgoprudny, Moscow region, 141700\\
Institute for Theoretical and Experimental Physics NRC ``Kurchatov Institute'', Moscow, 117218 Russia}

\author{A.\,A.~Nikolaev}
\email[]{aleksandr.nikolaev@swansea.ac.uk}
\affiliation{Department of Physics, College of Science, Swansea University, Swansea SA2 8PP, United Kingdom}

\date{\today}

\begin{abstract}
In this paper we study the properties of QCD at nonzero chiral density $\rho_5$, which is introduced through chiral chemical potential $\mu_5$. The study is performed within lattice simulation of QCD
with dynamical rooted staggered fermions. We first check that $\rho_5$ is generated at nonzero $\mu_5$ and in the chiral limit observe $\rho_5 \sim \Lambda_{QCD}^2 \mu_5$. We also test the possible connection between confinement and topological fluctuations. To this end, we measured the topological susceptibility $\chi_{\mbox{\footnotesize top}}$ and string tension $\sigma$ for various values of $\mu_5$. We observed that both string tension and chiral susceptibility grow with $\mu_5$ and there is a strong correlation between these quantities. We thus conclude that the chiral chemical potential enhances topological fluctuations and that these fluctuations can indeed be closely related to the strength of confinement.
\end{abstract}


\maketitle

\section{Introduction}

Quantum Chromodynamics (QCD) is believed to be the theory of strong interaction. While microscopic QCD Lagrangian is well known, the theory itself is extremely complicated and possesses a plenty of not fully understood nontrivial properties and phenomena.
The most well-known examples include color confinement and chiral symmetry breaking.

One of the possible ways to shed light on these phenomena and their mechanism is to investigate QCD or QCD-like theories under extreme conditions. These extreme conditions include finite temperature studies~\cite{Smilga:1996cm, DElia:2018fjp, Agasian:2016yck, Abramchuk:2018bco}, the influence of large magnetic field on QCD properties~\cite{Andersen:2014xxa, Kharzeev:2013jha, DElia:2015kpi, Orlovsky:2013aya, Andreichikov:2012xe,Braguta:2019yci}, QCD and QCD-like theories at finite baryon density~\cite{Muroya:2003qs,Fischer:2018sdj, Andreichikov:2017ncy,Braguta:2016cpw, Bornyakov:2017txe, Astrakhantsev:2018uzd, Boz:2018crd, Cotter:2012mb,Braguta:2019yci} and QCD at finite isospin density~\cite{Brandt:2017oyy,Brandt:2018wkp,Braguta:2019noz}.

Among others are the properties of QCD at nonzero chiral density. Systems with nonzero chiral density attract considerable attention because of unusual phenomena which take place in such systems. The renowned example of such phenomena is the chiral magnetic effect (CME)~\cite{Fukushima:2008xe, Vilenkin:1980fu}, the appearance of electric current in chiral medium along applied magnetic field.
Nonzero chiral density can be generated in heavy ion collisions either due to sphaleron transitions in quark-gluon plasma~\cite{Kharzeev:2007jp,Kotov:2018vyl} or due to the axial anomaly in parallel electric and magnetic fields~\cite{Ruggieri:2016lrn}. There are a lot of studies of QCD properties with chiral density which is introduced through nonzero chiral chemical potential~\cite{Gatto:2011wc, Chernodub:2011fr, Andrianov:2012dj, Andrianov:2013dta, Yu:2015hym, Khunjua:2017mkc, Khunjua:2017khh, Khunjua:2018sro, Andrianov:2017ely, Braguta:2015owi, Braguta:2015zta, Braguta:2016aov}.

One of the interesting questions which can be addressed is how the confinement and the chiral symmetry breaking in QCD are affected by nonzero chiral density. The influence of nonzero chiral chemical potential on the chiral symmetry breaking was considered in a number of theoretical papers~\cite{Gatto:2011wc, Chernodub:2011fr, Andrianov:2012dj, Andrianov:2013dta, Yu:2015hym, Braguta:2016aov,Andrianov:2017ely} as well as in the lattice studies~\cite{Braguta:2015owi, Braguta:2015zta}. Today it is clear that in any system the chiral chemical potential either creates or enhances the dynamical chiral symmetry breaking depending on the strength of interactions between constituents in the media. This phenomenon was called the chiral catalysis and the mechanism responsible for this phenomenon was first explained in~\cite{Braguta:2016aov}. The essence of this phenomenon is that nonzero chiral density generates additional fermionic states which take part in the formation of the chiral condensate. 

In this paper we mainly address three questions. First, we show that introduction of nonzero $\mu_5$ to the system Hamiltonian leads to generation of nonzero chiral density $\rho_5$. We study its dependence within lattice simulation of QCD and compare the observed behavior with the existing models such as ChPT and NJL. Second, we study the influence of chiral density on topological structure of QCD and show that topological susceptibility $\chi_{\mbox{\footnotesize top}}$ increases with chiral density $\rho_5$. Finally, we study the confinement in QCD with nonzero $\mu_5$ and its connection to the topology of QCD. The possible link between these phenomena was introduced in~\cite{Kharzeev:2015xsa,Kharzeev:2015ifa}. Namely, the authors suggested to modify the gluon propagator to have the form $G(p) = (p^2 + \chi_{\mbox{\footnotesize top}} / p^2)^{-1}$ due to
Veneziano ghosts tunnelling between different topological sectors of QCD. This form of gluon propagator implies maximum propagation range of order $\chi_{\mbox{\footnotesize top}}^{-1/4}$ and suggests enhancement of confinement with the growth of topological susceptibility. To check this connection, we study the string tension $\sigma$ between heavy quark and antiquark at nonzero $\mu_5$ and its correlation with topological susceptibility $\chi_{\mbox{\footnotesize top}}$. Our results support the idea that topological properties and confinement are tightly connected.

It is well known that introduction of baryon chemical potential leads to the sign problem in $SU(3)$ theory and spoils the LQCD simulations. On contrary, introduction of the chiral chemical potential does not lead to the sign problem~\cite{Fukushima:2008xe}, which allows us to carry out this study within lattice simulation of QCD. 

This paper is organized as follows. In the next section we discuss the chiral density generated by nonzero chiral chemical potential in QCD. In the section~\ref{sec:three} we describe the details of our lattice simulation. Our results are presented in the section~\ref{sec:four}. In the last section we discuss our results and draw the conclusions. In Appendix~\ref{app:A} we derive the chiral density for free "naive" fermions and study divergences in the chiral density.

\section{Nonzero chiral chemical potential in QCD}
\label{sec:two}

In this paper we are going to study the properties of QCD with nonzero chiral density $\rho_5=\bar \psi \gamma_4 \gamma_5 \psi$. It is well known that nonzero baryon density can be introduced to statistical system through modification of the Hamiltonian in the partition function $\hat H \to \hat H - \mu \int d^3 x \bar \psi \gamma_4 \psi$.~\footnote{In this paper we study QCD in thermodynamic equilibrium. So, instead of real time one has Euclidean time which is designated as a fourth component of four-vector. In particular, we use the following notation $\gamma_4=\gamma_0$} Similarly, one can modify the Hamiltonian by the term with chiral chemical potential $\mu_5$
\beq
\hat H \to \hat H - \mu_5 \int d^3 x \bar \psi \gamma_4 \gamma_5 \psi. 
\label{hamiltonian}
\eeq

We would like to stress that the chiral chemical potential is different to the baryon chemical potential since chiral density is not conserved. There are two operators resulting in the non-conservation of the chiral density 
\beq
\begin{split}
\frac d {dt} \int d^3 x \rho_5 = \frac {\alpha_s N_c} {4 \pi} \int d^3 x & F^a_{\mu \nu} {\tilde F}^a_{\mu \nu} +\\
&+2 m \int d^3 x \bar \psi \gamma_5 \psi.
\label{axial_anomaly}
\end{split}
\eeq

The first operator $\sim F^a_{\mu \nu} {\tilde F}^a_{\mu \nu}$ is the anomalous contribution due to quantum corrections. The second operator $\sim m \bar \psi \gamma_5 \psi$ results from the equation of motion for massive fermions. Note that chirality is not well-defined for massive fermions due to the possible spin flipping process. The dynamical fermion mass generation $\sim \Lambda_{QCD}$ due to the chiral symmetry breaking can significantly increase the effect of spin flipping. Thus, the physical meaning of modification (\ref{hamiltonian}) should be discussed more carefully. It is clear that the $\rho_5$ operator becomes the true chiral density only in the massless limit $\rho_5\bigl|_{m\to0}=(Q_R-Q_L)/V$. For massive quarks the meaning of the $\rho_5$ operator should be considered in more detail.

Chemical potential is usually introduced with respect to conserved charge. In our study we consider $\mu_5 \rho_5$ as the new term in the Hamiltonian and the conservation of $\rho_5$ is not required. We expect that the modification (\ref{hamiltonian}) leads to nonzero averaged value of the chiral density operator $\langle \rho_5 \rangle \neq 0$ even for nonzero quark mass. The situation with $\mu_5$ and $\rho_5$ is similar to the one with the fermion mass term $m \bar \psi \psi$. The conservation of the $\bar \psi \psi$ operator is not required and once this operator is introduced to the Hamiltonian it leads to the generation of nonzero condensate $\langle \bar \psi \psi \rangle \neq 0$. To show that it is very likely that non-zero $\mu_5$ will result in non-zero $\rho_5$ generation even at finite quark mass, let us consider various models of QCD.

First, in terms of fermionic spectrum, the modification of the Hamiltonian (\ref{hamiltonian}) modifies the dispersion relation $E^2(p)=(|\vec p| -  s \mu_5)^2 + m^2$~\cite{Gatto:2011wc}, where $s=\pm 1$ is the fermion helicity. For $\mu_5>0$ this implies that at fixed momentum $|\vec p|$ the fermion with helicity $s=+1$ has smaller energy than the one with $s=-1$. In thermodynamic equilibrium there will be a larger number of fermions with helicity $s=+1$ than that with $s=-1$. So one can expect that the modification (\ref{hamiltonian}) leads to nonzero helicity even at nonzero quark mass.

We proceed with the consideration of $SU(N_c)$ QCD with finite chemical potential in the large $N_c$ limit $N_c \to \infty$. At low temperature $T$ the chiral perturbation theory (ChPT)~\cite{Scherer:2002tk, Ecker:1998ai} can be applied. In the leading order in $1/N_c$ there is no contribution of the anomalous term. Modification (\ref{hamiltonian}) only adds the flavour singlet axial current $A_{\mu} = \mu_5 \delta_{\mu4} \hat 1$ and the modification of the partition function due to the introduction of this axial current within ChPT reads
\beq
Z(\mu_5)=Z_{QCD}\times \exp {( \beta V N_f f_{\pi}^2 \mu_5^2)}, \quad \beta = \frac 1 T.
\label{partition_function}
\eeq

From Eq.~(\ref{partition_function}) it is seen that nonzero $\mu_5$ leads to the additional constant factor in the QCD partition function, i.e. it is not related to dynamical degrees of freedom of the ChPT. This is because the ChPT accounts the chiral symmetry breaking in QCD but it does not provide its mechanism. In more complicated models~\cite{Witten:1980sp, DiVecchia:1980yfw} which consider the chiral symmetry breaking mechanism, the $\mu_5$ couples to the scalar $\sigma$ and $\eta'$ fields, which leads to enhancement of the chiral symmetry breaking with $\mu_5$ and the chiral catalysis phenomenon in QCD~\cite{Braguta:2016aov}. 

From (\ref{partition_function}) for two flavor QCD one has 
\beq
\langle \rho_5 \rangle= \frac 1  {\beta V} \frac {\partial \log Z(\mu_5)} {\partial \mu_5} = 4 f_{\pi}^2 \mu_5.
\label{rho5}
\eeq

So, one can see that the modification of the Hamiltonian (\ref{hamiltonian}) indeed leads to nonzero $\langle \rho_5 \rangle$ in the limit $N_c \to \infty$ even at nonzero quark mass. 

Similar study can be carried out in the Nambu-Jona-Lasinio (NJL) model~\cite{Klevansky:1992qe}, which successfully describes low energy phenomenology of QCD. Since the NJL model is usually studied within the saddle point approximation, a lot of results are obtained within the $N_c \to \infty$ assumption. Within the NJL model, the chiral symmetry breaking leads to the generation of the dynamical quark mass $m\sim \Lambda_{QCD}$. The calculation of the chiral density with the quark mass $m\sim \Lambda_{QCD}$ gives $\langle \rho_5 \rangle \sim \Lambda_{QCD}^2 \mu_5$ (see Appendix~\ref{app:A}). This is another argument in favor of the hypothesis that non-zero chiral density can be generated at finite mass $\langle \rho_5 \rangle \sim \Lambda_{QCD}^2 \mu_5 \neq 0$.

On the one hand the approximation $N_c \to \infty$ works quite well for real QCD. So, one might expect that $\langle \rho_5 \rangle \sim \Lambda_{QCD}^2 \mu_5 \neq 0$ for $N_c=3$. However, the anomaly contribution which appears in higher orders $\sim 1/N_c$--corrections can modify the $N_c \to \infty$ result for the chiral density. To clarify the $N_c = 3$ behavior in Section~\ref{sec:four} we conduct lattice study of chiral density $\rho_5$ at nonzero chiral chemical potential. 

\section{Lattice setup}
\label{sec:three}

In this paper we are going to study QCD with two flavours and nonzero chiral chemical potential. 
To this end we perform lattice simulations with the $SU(3)$ gauge group and employed the tree level improved Symanzik gauge action~\cite{Weisz:1982zw,Curci:1983an}. For the fermionic part of the action 
we used staggered fermions with the action~\cite{Braguta:2015zta}
\begin{equation}\begin{split}
S_f&=ma\sum_x {\bar \psi_x}\psi_x + \\
& +\frac12\sum_{x\mu} 
   \eta_{\mu}(x)({\bar \psi_{x+\mu}U_{\mu}(x)}\psi_x
   -{\bar \psi_x}U^{\dag}_{\mu}(x) \psi_{x+\mu}) + \\
   & + \frac12\mu_5a\sum_x s(x)({\bar \psi}_{x+\delta}{\bar U}_{x+\delta, x}\psi_x
   -{\bar \psi}_{x}{\bar U}_{x+\delta, x}^{\dag}\psi_{x+\delta}),
\label{eq:staggeredaction}
\end{split}\end{equation}
where the $\eta_{\mu}(x)$ are the standard staggered phase factors: 
$\eta_1(x)=1,\eta_{\mu}(x)=(-1)^{x_1+\ldots+x_{\mu-1}}$ for $\mu=2,3,4$
\footnote{It is important to note that staggered fermions generate 
the correct non-abelian chiral anomaly \cite{Coste:1986qr, Jolicoeur:1987rj} }.
The lattice spacing is denoted by $a$, the bare fermion mass by $m$,
and $\mu_5$ is the chiral chemical potential.
In the chirality breaking term $s(x)=(-1)^{x_2}$, $\delta=(1,1,1,0)$ 
represents a shift to the diagonally opposite site in a spatial $2^3$ 
elementary cube. The combination of three links connecting sites $x$ and 
$x+\delta$,
\begin{equation}\begin{split}
{\bar U}_{x+\delta,x}=\frac16\sum\limits_{i,j,k=
\text{perm}(1,2,3)}U_i(x+e_j+e_k)U_j(x+e_k)U_k(x) 
\end{split}\end{equation}
is symmetrized over the $6$ shortest paths between these sites. In the 
partition function, after integrating out fermions, one 
obtains the corresponding fermionic determinant. In order to obtain two flavours in the 
continuum limit we apply the rooting procedure.

In the continuum limit and after rooting procedure our lattice action can be rewritten in the
Dirac spinor-flavor basis \cite{KlubergStern:1983dg,MontvayMuenster:2000} as 
follows  
\begin{equation}\begin{split}
S_f \to & S^{(cont)}_f = \\
&=\int d^4x \sum_{i=1}^2 \bar{q_i}
(\partial_{\mu}\gamma_{\mu} +igA_{\mu}\gamma_{\mu}+m+\mu_5\gamma_5\gamma_4)q_i.
\end{split}\end{equation}

We would like to emphasize that the chiral chemical potential introduced 
in Eq.~(\ref{eq:staggeredaction}) corresponds to the taste-singlet operator 
$\gamma_5\gamma_4\otimes\boldsymbol{1}$ in the continuum limit.

It should be also noted here that the baryonic chemical potential~\cite{Hasenfratz:1983ba} and the chiral chemical potential as in~\cite{Yamamoto:2011ks}, are introduced to the action as the modification of the temporal links by the corresponding exponential factors in order to eliminate chemical-potential 
dependent quadratic divergences. For staggered fermions with the baryonic chemical potential this modification 
can be performed. However, in the case of $\mu_5$ this method would lead to a highly 
non-local action~\cite{Yamamoto:2011ks}. 
Therefore, we introduce $\mu_5$ in Eq.~(\ref{eq:staggeredaction}) in the additive way similarly to the mass term. 
It is known that the additive introduction of the chemical potential might lead to additional divergences in observables. In this paper we perform lattice measurement of chiral density and gluonic observables: the topological charge, the topological susceptibility and the string tension. In what follows we account ultraviolet divergences in the chiral density. We also believe that there are no additional divergences due to chiral chemical potential in gluon observables, because the chiral chemical potential term can be considered as some vertex with coupling constant of dimension of energy. It is known that the inclusion of such vertex to Feynman diagrams reduces the power of ultraviolet divergences. Since the fermion loops in QCD diverge as powers of $\log a$, the chiral chemical potential does not give rise to additional divergences. The ultraviolet divergences in QCD with chiral chemical potential are also discussed in~\cite{Braguta:2015zta, Braguta:2015owi}. 

The physical lattice spacing $a$ was determined from setting Sommer parameter $r_0$~\cite{Sommer:1993ce} to its physical values $r_0 = 0.468(4)$ fm~\cite{Bazavov:2011nk}. Simulation for scale setting were performed with the lattice size $24^4$, $\mu_5 = 0$ and fixed $ma = 0.01$. Since the Sommer scale very mildly depends on the quark mass~\cite{Sommer:2014mea}, the physical units are almost independent from the quark mass. Notice also that as was shown in papers \cite{Braguta:2015zta, Braguta:2015owi} nonzero $\mu_5$ does not affect to the scale 
setting procedure.

\begin{widetext}
\begin{center}
\begin{table}[h!]
    \centering
    \begin{tabular}{c|c|c|c|c}
     $\beta$ & $a$, fm & $L^4$ & $ma$ & $\mu_5a$ \\
     \hline
     3.9 & 0.128(3) & $14^4$ & 0.0148, 0.0296, 0.0445 & 0.0, 0.152, 0.304, 0.365, 0.487, 0.609\\
     4.0 & 0.1054(11) & $16^4$ & 0.01, 0.02, 0.03 & 0.0, 0.125, 0.25, 0.30, 0.40, 0.50\\
     4.1 & 0.0856(14) & $20^4$ & 0.00658, 0.01316, 0.1974 & 0.0, 0.1015, 0.2030, 0.2436, 0.3248, 0.4060\\
     \hline
    \end{tabular}
    \caption{Lattice parameters used in the simulations}
    \label{tab:parameters}
\end{table}
\end{center}
\end{widetext}

In the calculation we employed three different lattices with different lattice spacings to keep the physical volume fixed at approximately $1.7\,\mbox{fm}^3$: $14^4$ with $a = 0.128(3)\,\mbox{fm}$ ($\beta = 3.9$), $16^4$ with $a = 0.1054(11)\,\mbox{fm}$ ($\beta = 4.0$) and $20^4$ with $a = 0.0856(14)\,\mbox{fm}$ ($\beta = 4.1$). To investigate chiral properties for each of the listed lattices three values of pion mass were considered: $m_\pi = 563,\,762,\,910\,\mbox{MeV}$. We summarize lattice parameters of the simulations in Tab.~\ref{tab:parameters}. We note that the simulations performed in this paper indicate that the required simulation time grows with the chiral chemical potential. Lattice simulations at the largest values of chiral chemical potential are numerically very expensive.

\section{Results of the calculation}
\label{sec:four}

\subsection{The chiral density}

\begin{figure}
    \centering
    \includegraphics[scale = 0.6]{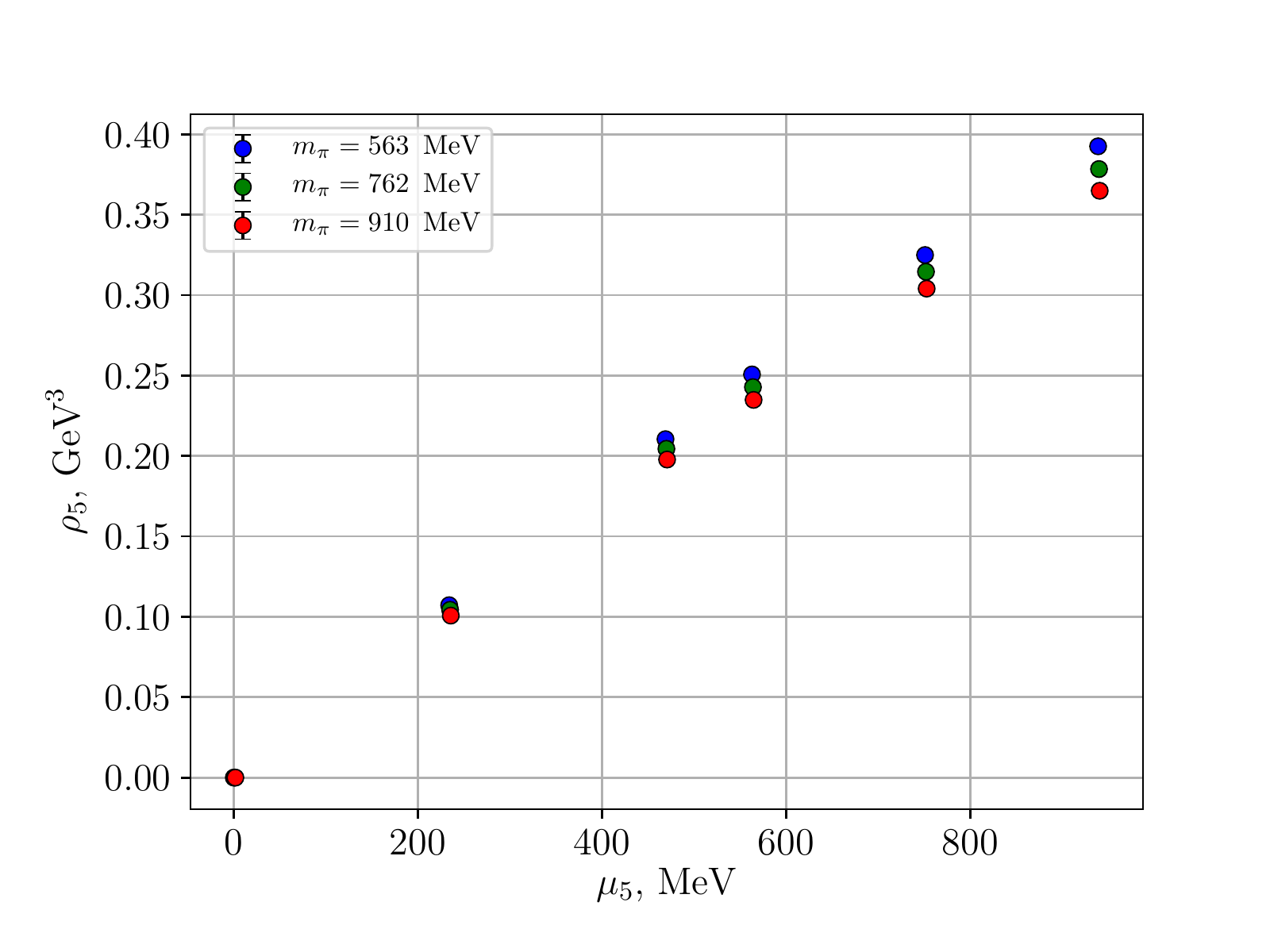}
    \caption{The chiral densities as a function of the chiral chemical potential for different pion masses and $a=0.105\,\mbox{fm}$.}
    \label{fig:chiral_density}
\end{figure}

In this section we perform lattice measurement of the chiral density $\rho_5$ for
all lattice spacings and pion masses under study. The chiral densities as a function of the 
chiral chemical potential for different pion masses and $a=0.105\,\mbox{fm}$ are shown in Fig.~\ref{fig:chiral_density}. The chiral densities for other lattice spacings look 
similar. For this reason we do not show them. From Fig.~\ref{fig:chiral_density} one sees that the data are well described by the linear dependence. It turns out that the coefficient of this linear dependence can be mostly attributed to the ultraviolet divergence in $\rho_5$. However, our data are rather accurate. Typical uncertainty of the calculation is $\sim 0.1 \%$, for this reason we can extract the sub-leading terms on the background of leading ultraviolet divergence. 

To proceed we need to know the structure of the divergences in $\rho_5$. In Appendix~\ref{app:A} the study of the ultraviolet divergences in $\rho_5$ for free "naive" fermions is presented. In particular, it is shown that there are two ultraviolet divergences in the term linear in $\mu_5$. The leading divergence is quadratic and the next-to-leading divergence is logarithmic. Additionally, the linear in $\mu_5$ term contains finite contribution. Finally higher terms in $\mu_5$ expansion do not contain ultraviolet divergences. 

In this paper we are going to use the following anzats for $\rho_5$ which accounts for the results obtained in Appendix~\ref{app:A}:
\beq
\begin{split}
a^3 \rho_5 = E (a \mu_5)^3 + (A + a^2 B + C_1 (m a)^2 + \\ +D (ma)^2 \log (ma)^2 + F a^2 (ma)^2 + X a^4) \times (a \mu_5).
\label{eq:rho5_formula}
\end{split}
\eeq

This fit gives decent description of the data $\chi^2 / \mbox{ndof} \sim 3$. Since the measurements of $\rho_5$ are quite accurate (at some points the error is only 0.05\,\%), we are able to fix all the parameters with the error of not worse than 15\,\%. Removing of any of the terms in~(\ref{eq:rho5_formula}) leads to significant growth of $\chi^2 / \mbox{ndof}.$ However, adding higher powers of $ma$ and $a$ to the fit does not improve the quality.

It is important to notice that the coefficient $B$ from~(\ref{eq:rho5_formula}) is non-zero and $B = (340(10)\,\mbox{MeV})^2$. This coefficient parameterizes 
the chiral density in the continuum and in the chiral limits. For this reason we can 
state that $B \sim \Lambda_{QCD}^2$ or $\rho_5 \sim \Lambda_{QCD}^2 \mu_5$. 
Notice, however, that it is not possible to write exactly $\rho_5 = B\mu_5$ since the multiplicative renormalization of $\rho_5$ might be important but goes beyond the scope of this paper. To summarize, the results of this section allow us to state that finite $\mu_5$ generates nonzero chiral density $\rho_5 \sim \Lambda_{QCD}^2 \mu_5 + O(\mu_5^3)$.

\subsection{The topological charge and topological susceptibility}
\label{subsec:chiT}

\begin{figure}[h!]
    \includegraphics[scale = 0.5]{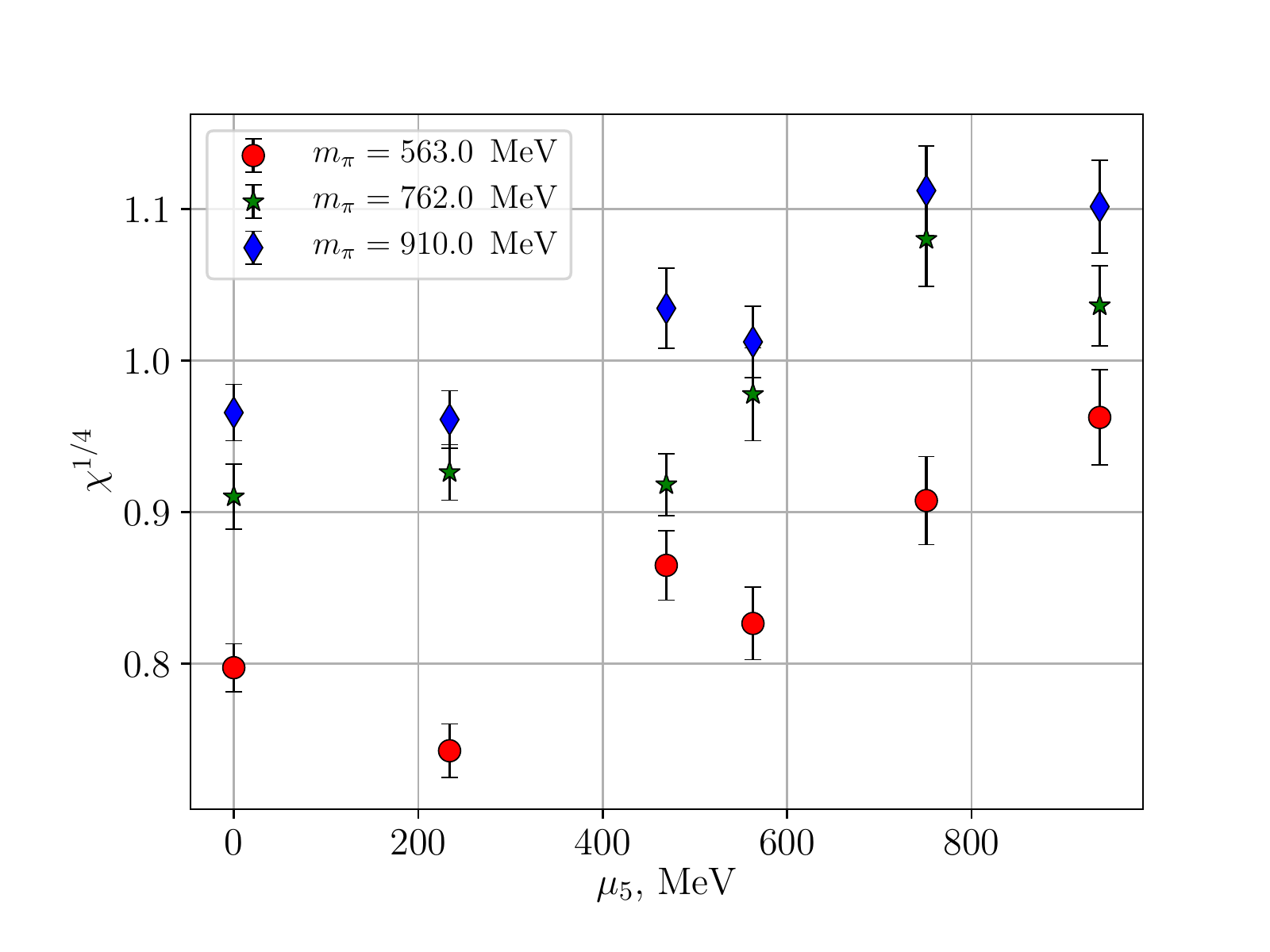}
\caption{The topological susceptibility as the function of the chiral chemical potential for different pion masses in the continuum limit performed using the non-singlet pion mass correction of $\chi^{1/4}(a)$ procedure described in App.~\ref{app:B}.}
    \label{fig:chi}
\end{figure}

Our next task is to study how nonzero chiral chemical potential influences the topological properties of QCD. To this end we measure the topological charge $\left\langle Q \right\rangle$ and the topological susceptibility $\left\langle Q^2 \right\rangle$ for different values of the chiral chemical potential under study. 

Our measurement of the topological charge and the topological susceptibility mainly follows~\cite{Bonati:2014tqa}. We smoothen each configuration using the Gradient Flow~\cite{Luscher:2009eq,Luscher:2010iy}. Topological charge is measured on the smoothened configurations
\begin{equation}
    Q_L=-\frac{1}{512\pi^2}\sum_x\sum_{\mu\nu\rho\sigma=\pm1}^{\pm4}\tilde{\epsilon}_{\mu\nu\rho\sigma}\tr U_{\mu\nu}(x)U_{\rho\sigma}(x)\,,
    \end{equation}
where $U_{\mu\nu}(x)$ is the plaquette at the point $x$ in directions $\mu$ and $\nu$. In order to reduce the lattice artifacts we used the following estimators of the topological charge $Q$:
\begin{equation}
    Q=\roundcustom\left(\alpha Q_L\right),
\end{equation}
where $\roundcustom$ gives the closest integer to its argument and the factor $\alpha$ is chosen in such a way that it minimizes 
\begin{equation}
    \langle \left(\alpha Q_L-\roundcustom\left(\alpha Q_L\right)\right)^2 \rangle\,.
\end{equation}

\begin{figure}[h!]
    \includegraphics[scale = 0.5]{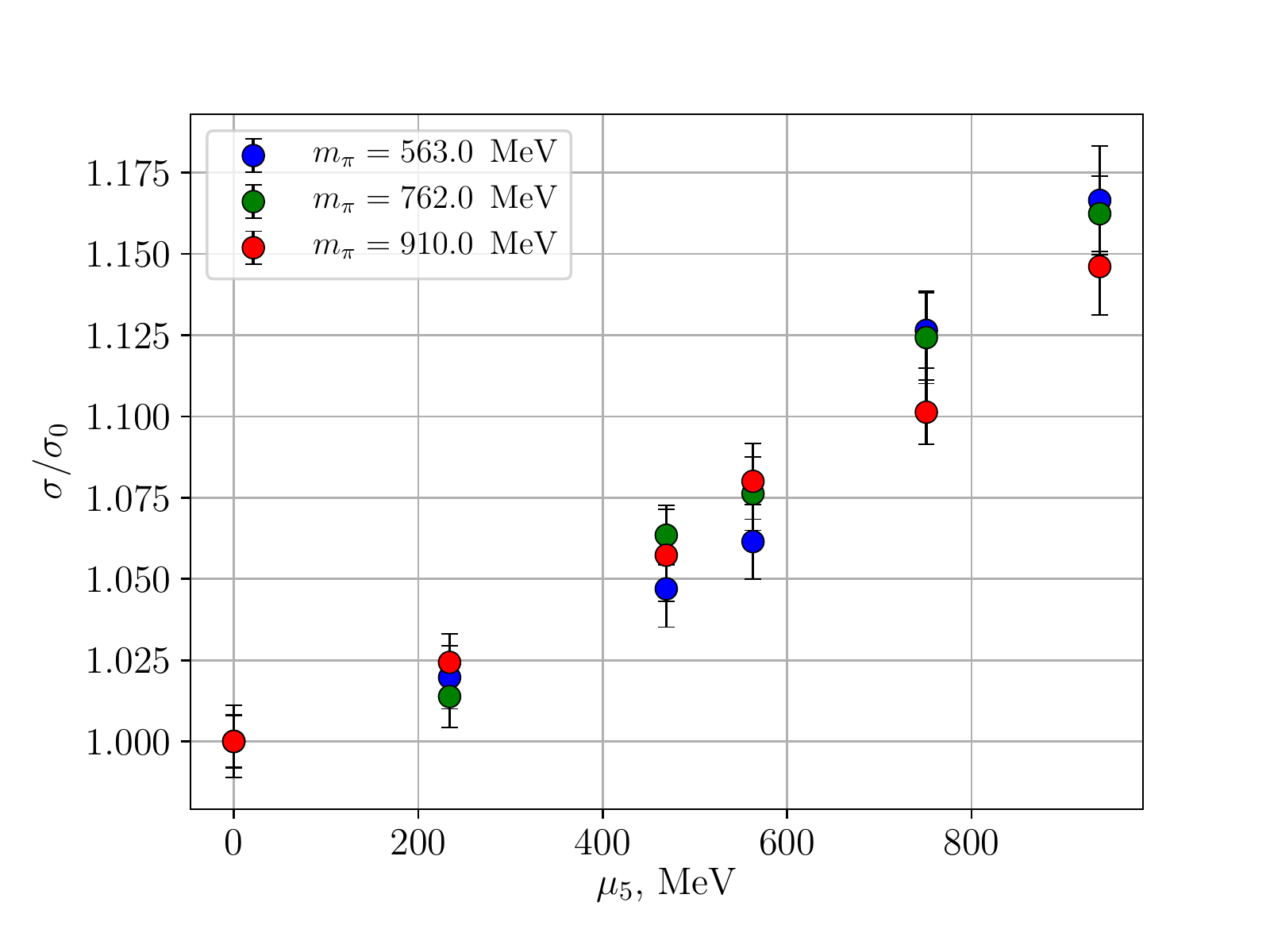}
\caption{The ratio of the string tension $\sigma$ to the string tension at zero chiral chemical potential $\sigma_0$. Points for different pion masses are slightly shifted in horizontal axis for better visibility.}
    \label{fig:s_s_0}
\end{figure}

In other words, we rescale our definition of the topological charge $Q_L$ so that its peaks become closer to integer values and then round the result to this integer value. The topological susceptibility is then defined as
\begin{equation}
    \chi_{\mbox{\footnotesize top}}=\frac{\langle Q^2 \rangle}{V_4}\,,
\end{equation}
where $V_4$ is the four-dimensional volume of the lattice. We have found that for Gradient Flow times $t/a^2>3.0$ the dependence of the topological susceptibility $\chi_{\mbox{\footnotesize top}}$ on the value of Gradient Flow time exhibits a plateau with almost no dependence on the value of $t$. The value at this plateau was taken as a final estimation for the topological susceptibility $\chi_{\mbox{\footnotesize top}}$. In the Appendix~\ref{app:B} we show that discretization errors in the topological susceptibility are under control.

Our results for the topological properties of QCD are the following.
The topological charge is zero within the uncertainty of the calculations 
for all pion masses, lattice spacings and chiral chemical potentials under study.

In Fig.~\ref{fig:chi} we show the topological susceptibility as the function of chiral chemical potential for different pion masses in the continuum limit performed according to procedure described in Appendix~\ref{app:B}. One sees that the chiral chemical potential indeed enhances the topological fluctuations in QCD for all pion masses. For this reason, we believe that the chiral chemical potential enhances the topological fluctuations in QCD. 

A possible explanation of this fact is the following. As we know nonzero chiral chemical potential 
leads to generation of nonzero chiral charge in the system with some average value $Q_5$. Due to the anomaly this 
chiral charge can annihilate to gluon configurations with nonzero Chern-Simons number,
which is compensated by an inverse process: creation of the chiral charge from 
gluon background with Chern-Simons number. In the thermodynamic equilibrium these processes compensate each other leading to some fixed average value of the chiral density. 
Notice also that both processes result from the chiral anomaly. Further let us consider the process of 
annihilation of the chiral charge as a number of elementary processes in which one quark and one 
antiquark annihilate to gluon configuration with nonzero Chern-Simons number. It is 
reasonable to assume that the larger the chiral charge the larger the number of elementary 
annihilations per time unit in the system. In other words the larger the chiral charge $Q_5$
the larger average $\langle d Q_5 / dt \rangle_{annihilation}$ for the annihilation processes. Notice 
that is completely compensated by the inverse process leading to the total $\langle d Q_5 / dt \rangle=0$
From this picture one can expect that the larger the 
chiral charge the larger the topological fluctuations in the system under investigation.
Our results imply that $\mu_5$ is the parameter which allows influencing the topological sector of QCD through the anomaly equation.

\subsection{The string tension}
In order to study how nonzero chiral density influences the confinement properties of QCD we calculated the interaction potential of static charges through the measurement of Wilson loops. 
To obtain reasonable signal-to-noise ratio for Wilson loops the smearing techniques were employed. One step of the hypercubic blocking~\cite{Hasenfratz:2001hp} with parameters $\alpha = (1.0,\,1.0,\,0.5)$~\cite{DellaMorte:2005nwx} was performed for the temporal links only, followed by 24 steps of the APE smearing~\cite{Albanese:1987ds} with $\alpha_{APE} = 0.165$.

The quark-antiquark interaction potential is related to Wilson loops as
\begin{equation}\label{eq:W_plateau}
V(R) = \lim_{t \rightarrow \infty} \text{log} \Bigl[ \frac{\left\langle W(R,t) \right\rangle}{\left\langle W(R,t+1) \right\rangle} \Bigr]\,.
\end{equation}
This logarithm exhibits a clear plateau at large times $t/a\in[5; 9]$, its height was extracted as $V(R)$.

String tension $\sigma$ was obtained from fitting of the potential in the range $R \in [3.5 a ; L_s / 2]$ by the Cornell fit
\begin{equation}\label{eq:Cornell_fit}
V(R) = A - \alpha / R + \sigma R\,.
\end{equation}
This fit provides $\chi^2 / dof \lesssim 1$ for all values of chiral chemical potentials. To estimate systematic uncertainty the left fitting range was varied in the interval $[3 a; 4 a]$ and the produced small change of $\sim 0.5\%$ in the string tension was added to the statistical error. Change of the right boundary of $R$ in the fit does not alter the results in a noticeable way. Statistical errors for the fit parameters were estimated with the jackknife method.

It is worth to note, that the Wilson loop corresponds to an operator, that creates the static color sources and a string between them, and this operator has a small overlap with the state, corresponding to a broken string~\cite{Bali:2005fu}, thus the string breaking phenomenon can not be observed from the Wilson loops. On the other hand, for finite lattice due to the p.b.c. in spatial directions the maximal achievable separation between $q$ and $\bar q$ is $L_s/2$, which in our case corresponds to 0.85 fm ($L_s \approx 1.7$ fm, see Table~\ref{tab:parameters}). The string breaking for the physical pion mass appears near 1 fm, and in our study due to the heavier pions this should occur ever at larger $q \bar q$ separation. Thus extraction of quark-antiquark interaction potentials from Wilson loops does not lead to any problems as far as we investigate $V(R)$ at distances, which are smaller than the string breaking distance. The same argument applies to the choice of Cornell potential for $V(R)$ fitting.

We do not observe any significant dependence of the string tension on the lattice spacing $a$. Thus, we perform the constant fit of the string tension versus $a$ to average over different lattice spacings.  The quality of such fit is good, $\chi^2 / \mbox{ndof} < 1$. The ratio of the string tension $\sigma$ (extrapolated to continuum in the above described way) to the string tension at zero chiral chemical potential $\sigma_0$ is presented in Fig.~\ref{fig:s_s_0}.
It is seen from Fig.~\ref{fig:s_s_0} that the string tension rises with the chiral chemical potential i.e. with the chiral density.

\subsection{Topological fluctuations and confinement}
The phenomenon of the QCD confinement is not well understood on the present day. However, papers~\cite{Kharzeev:2015xsa, Kharzeev:2015ifa} have established a possible link between confinement properties and QCD topology. In their setup, gluon propagator is modified by the interaction with Veneziano ghosts tunneling between different topological sectors. The gluon propagator then reads $G(p) = (p^2 + \chi_{\mbox{\footnotesize top}} / p^2)^{-1}$, where $\chi_{\mbox{\footnotesize top}}$ is the topological susceptibility. The propagator has only complex poles $p^2 = \pm i \chi_{\mbox{\footnotesize top}}^{1/2}$, thus gluons cannot propagate as free particles. The typical range of gluon propagation decreases as $\chi_{\mbox{\footnotesize top}}^{-1/4}$ with the growth of topological susceptibility. As one can observe from (section~\ref{subsec:chiT}) the topological susceptibility is enhanced by $\mu_5$. Thus, the confining properties, namely the string tension should also be enhanced by $\mu_5$ and this is exactly our observation. 

\section{Conclusion and discussion}

In this paper we studied the properties of QCD at nonzero chiral density $\rho_5$, which is introduced through the chiral chemical potential $\mu_5$. Contrary to the baryon chemical potential introduction of the chiral chemical potential does not lead to the sign problem. For this reason our study of QCD with nonzero chemical potential can be performed within lattice simulation. In the simulations we employed the tree level improved Symanzik gauge action and rooted staggered fermions which in the continuum limit correspond to $N_f=2$ dynamical quarks.

In the calculation we employed three different lattices with different lattice spacings to keep the physical volume fixed at approximately $1.7\,\mbox{fm}^3$: $14^4$ with $a = 0.128(3)\,\mbox{fm}$ ($\beta = 3.9$), $16^4$ with $a = 0.1054(11)\,\mbox{fm}$ ($\beta = 4.0$) and $20^4$ with $a = 0.0856(14)\,\mbox{fm}$ ($\beta = 4.1$). To investigate the chiral properties for each of the listed lattices three values of pion mass were considered, $m_\pi = 563,\,762,\,910\,\mbox{MeV}$.

The first observable considered in this paper is the chiral density. We found that nonzero chiral chemical potential leads to generation of nonzero chiral density in QCD. Our lattice results support ChPT formula for the chiral density $\rho_5 \sim \Lambda_{QCD}^2 \mu_5$. 

The next question is the influence of nonzero chiral chemical potential on the topological properties of QCD. To address this question we measured the topological charge and the topological susceptibility for various values of $\mu_5$. We found that the topological charge is zero for all 
values of the chiral chemical potential under investigation. On the contrary, we found that the topological susceptibility rises with $\mu_5$. So we conclude that the chiral chemical potential or chiral density enhances the topological fluctuations in QCD.  

We believe that this observation can be understood as follows. Note that the chiral density enters the chiral anomaly which implies that the change of $\rho_5$ can generate nonzero topological charge. For this reason one can expect that the fluctuations of the chiral density lead to the fluctuations in the topological charge due to the axial anomaly. Larger chiral density generated by larger $\mu_5$ leads to larger fluctuations of the chiral density and, due to the anomaly, to larger topological fluctuations in QCD.  

The last observable studied in this paper is the string tension. We calculated the static potential from Wilson loops and determined the string tension for all values of chiral chemical potentials at lattice parameters studied. We found that the string tension rises with rising chiral chemical potential. 

It would be interesting to understand the mechanism how confinement in QCD is enhanced by nonzero chiral chemical potential. One possible explanation can be based on the results of~\cite{Kharzeev:2015xsa, Kharzeev:2015ifa}, where the authors considered the gluon propagator, modified due to Veneziano ghosts tunneling between different topological sectors, making gluons confined at typical distances $\sim \chi_{\mbox{\footnotesize top}}^{-1/4}$ where $\chi_{\mbox{\footnotesize top}}$ is the topological susceptibility. As one can observe from (section~\ref{subsec:chiT}) the topological susceptibility is enhanced by $\mu_5$. Thus, the confining properties, namely the string tension should also be enhanced by $\mu_5$ and this is exactly our observation.

Another possible explanation is that the gluon fields generated in the system due to fluctuations of $\rho_5$ might have nontrivial properties which give rise to the confinement. In particular, if the gluon fields are self-dual due to the $\rho_5$ fluctuations they might enhance the confinement~\cite{Leutwyler:1980ev, Efimov:1995uz, Nedelko:2014sla, Nedelko:2016gdk}. Unfortunately, quite large uncertainties of the calculation do not allow us to draw any strong conclusion about the origin of confinement enhancement with $\mu_5$. This question including the mechanism of self-dual gluon fields is the subject for further research.

\acknowledgments

V.\,V.\,B. acknowledges the support from the BASIS foundation. The work of N.\,Yu.\,A. and A.\,Yu.\,K., which consisted of generation of configurations and measurement of topological susceptibility, string tension and chiral density, was supported by grant from the Russian Science Foundation (project number 18-72-00055). A.\,A.\,N. acknowledges the support from STFC via grant ST/P00055X/1. This work has been carried out using computing resources of the federal collective usage center Complex for Simulation and Data Processing for Mega-science Facilities at NRC ``Kurchatov Institute'',~\url{http://ckp.nrcki.ru/}. In addition, the authors used the equipment of the shared research facilities of HPC computing resources at Lomonosov Moscow State University, the cluster of the Institute for Theoretical and Experimental Physics and the supercomputer of Joint Institute for Nuclear Research ``Govorun''.

\appendix

\section{Ultraviolet divergences in the chiral density for free "naive" fermions}
\label{app:A}

To get an idea about the ultraviolet divergences in the chiral density at nonzero chiral chemical potential in this section we are going to derive the chiral density for free "naive" fermions. 
The fermion propagator including the chiral chemical potential for "naive" 
lattice fermions can be written in the following form
\begin{widetext} 
\beq
S^{\alpha \beta}(x,y) &=& \frac {\delta^{\alpha \beta}} {L_t L_s^3} \sum_{ \{p\} } \sum_s  e^{i p (x-y)} \frac {-i \sum_{\mu} \gamma_{\mu} \sin (p_{\mu}) + ma + (\mu_5a) \gamma_4 \gamma_5 }
 {\sin^2 (p_4) + (|p|-s (\mu_5a))^2 +(ma)^2} \times P(s),
\\ \nonumber
P(s) &=& \frac 1 2 \biggl ( 1 -  i s \sum_i \frac {\gamma_{i} \sin (p_{i}) } {|p|} \gamma_0 \gamma_5  \biggr ),~~~i=1,2,3,
\\ \nonumber
|p|^2 &=& \sin^2 (p_1) + \sin^2 (p_2) + \sin^2 (p_3),
\\ \nonumber
p_i &=& \frac {2 \pi} {L_s} n_i,~~ i=1,2,3,~~ n_i=0,...,L_s-1, \\ \nonumber
p_4 &=& \frac {2 \pi} {L_t} n_4 + \frac {\pi} {L_t},~~~ n_4=0,...,L_t-1.
\eeq
\end{widetext}
Here $m$ and $\mu_5$ are mass and chiral chemical potential in physical units,
$\alpha, \beta$ are color indices, the sum is taken over all possible values 
of $(n_1, n_2, n_3, n_4)$, $s= \pm 1$ and $a$ is a lattice spacing. 

In the limit $L_s, L_t \to \infty$ the chiral density in lattice units for two fermion flavours can be written as
\begin{widetext}
\beq
\langle \bar \psi \gamma_4 \gamma_5 \psi \rangle_{lat} = - \frac{3 N_f}{16} Sp\bigl [ \gamma_4 \gamma_5 S(x,x) \bigr ] = -\frac 3 4 \sum_{s=\pm 1} \int \frac {d^4 p} {(2\pi)^4} \frac {s |p| - \mu_5a } {\sin^2 p_4 + (|p|-s(\mu_5a))^2 + (ma)^2 }
\label{chiral_density}
\eeq
\end{widetext}
It should be noted here that the factor $3$ in the first equality is due to the sum over the fermion colors. 
Now let us expand the chiral density (\ref{chiral_density}) in powers of the chiral chemical potential.
It turns out that it is sufficient to keep only two terms: $\sim \mu_5$ and $\sim \mu_5^3$. 
Higher order terms in this expansion do not contain ultraviolet divergences. 
 The calculation of the integrals which appear in this expansion is rather cumbersome but straightforward.
 For this reason we don't show the details of the calculation. We would like only to mention
 that the integrals which appear in the expansion of (\ref{chiral_density}) in $\mu_5$ can be found in 
  \cite{Capitani:2002mp}. The resulting expression for the chiral density $ \rho_5$ in physical units can be written in the following form
\beq
\label{chiral_density_res}
&&\rho_5 = \frac 1 {a^3} \langle \bar \psi \gamma_4 \gamma_5 \psi \rangle_{lat} = 
 {\mu_5}  J_1 + {\mu_5^3}  J_2 + O(\mu_5^5) \\ \nonumber
&&J_1= -0.464800 \frac 1 {a^2} - \frac 3 {\pi^2} m^2 \log (ma)^2 + 0.807241 m^2      \\ \nonumber
&&J_2= 0.242419
\eeq
Now few comments are in order. 
\begin{itemize}

\item From equation (\ref{chiral_density_res}) we notice that there are two divergences in the linear in the $\mu_5$ term. The leading divergence is quadratic and the next-to-leading divergence is logarithmic.

\item In addition to the divergences the linear in the $\mu_5$ term contains finite contribution which is proportional to the fermion mass in the second power $\rho_5 \sim m^2 \mu_5$. Now recall that in Nambu-Jona-Lasinio model~\cite{Klevansky:1992qe}, which successfully describes low energy phenomenology of QCD, the chiral symmetry breaking leads to generation of the dynamical fermion mass $m \sim \Lambda_{QCD}$. For this reason one can expect that due to chiral symmetry breaking in QCD the renormalized $\rho_5 \sim \Lambda_{QCD}^2 \mu_5$. ChPT confirms this statement (see section~\ref{sec:two}).

\item Notice also that the logarithmic ultraviolet divergence is also possible in the $\mu_5^3$ term. However, final result does not contain the logarithmic ultraviolet divergence.

\item  If one takes the chiral limit in formula (\ref{eq:staggeredaction}), it is possible to get rid of logarithmic divergence as well as final term proportional to $m^2$. Unfortunately it is not possible to get rid of the $1/a^2$ divergence which results from the additive way of introducing the chiral chemical potential. In addition to the divergence the additive chemical potential modifies the coefficient in front of the $\mu_5^3$ contribution. For free chiral fermions this coefficient is determined by Fermi distribution and 
for three colors and two flavours it is 
$\rho_5(\mu_5)=2/\pi^2 \cdot \mu_5^3\simeq 0.202642 \cdot \mu_5^3$ (\cite{Fukushima:2008xe}). Comparing this value with the $J_2$ in formula (\ref{chiral_density_res}) it is seen that lattice artificial contribution is rather small but it is present. In this paper we concentrate on the linear in $\mu_5$ term, thus the fact that the coefficient of the $\mu_5^3$ term is modified by lattice artifacts does not affect the results of this paper. 

\end{itemize}

\section{Topological susceptibility in the continuum limit}
\label{app:B}
It is known that topological susceptibility $\chi$ suffers from large discretization errors. To check the dependency of our data on a finite lattice step, we performed continuum extrapolation for $\chi$, using two various procedures, described in~\cite{billeter2004topological,bonati2016axion}.
In the case of zero chiral chemical potential $\mu_5=0$ we compare our results with the predictions of ChPT.

First of all, in~\cite{bonati2016axion} it is noted that the dominant source of lattice artefacts in $\chi^{1/4}(a)$ is the chiral symmetry breaking present at finite lattice spacing in the staggered discretization. The dependence of $\chi^{1/4}(a)$ on the lattice spacing can be significantly reduced if instead of $\chi^{1/4}(a)$ the quantity 
\begin{equation}
    \chi^{1/4}_{tc}(a) = \frac{m_{\pi}}{m_{ngb}(a)} \chi^{1/4}(a) 
    \label{eq:taste_breaking}
\end{equation}
is considered. Here $m_{ngb}(a)$ is the mass of one of the non-Goldstone pions, i.e. of a state that becomes massless in the chiral limit only if the continuum limit is taken. Clearly, $m_{ngb} \to m_{\pi}$ as $a \to 0$, so in the continuum limit $\chi^{1/4}_{tc}(a) \to \chi^{1/4}(a).$ Following~\cite{bonati2016axion}, as the $m_{ngb}(a)$ we used the state with the taste structure $\gamma_i \gamma_{\mu}$, which mass is close to the root mean square of all other taste masses. 

\begin{figure}[h!]
    \centering
    \includegraphics[scale=0.55]{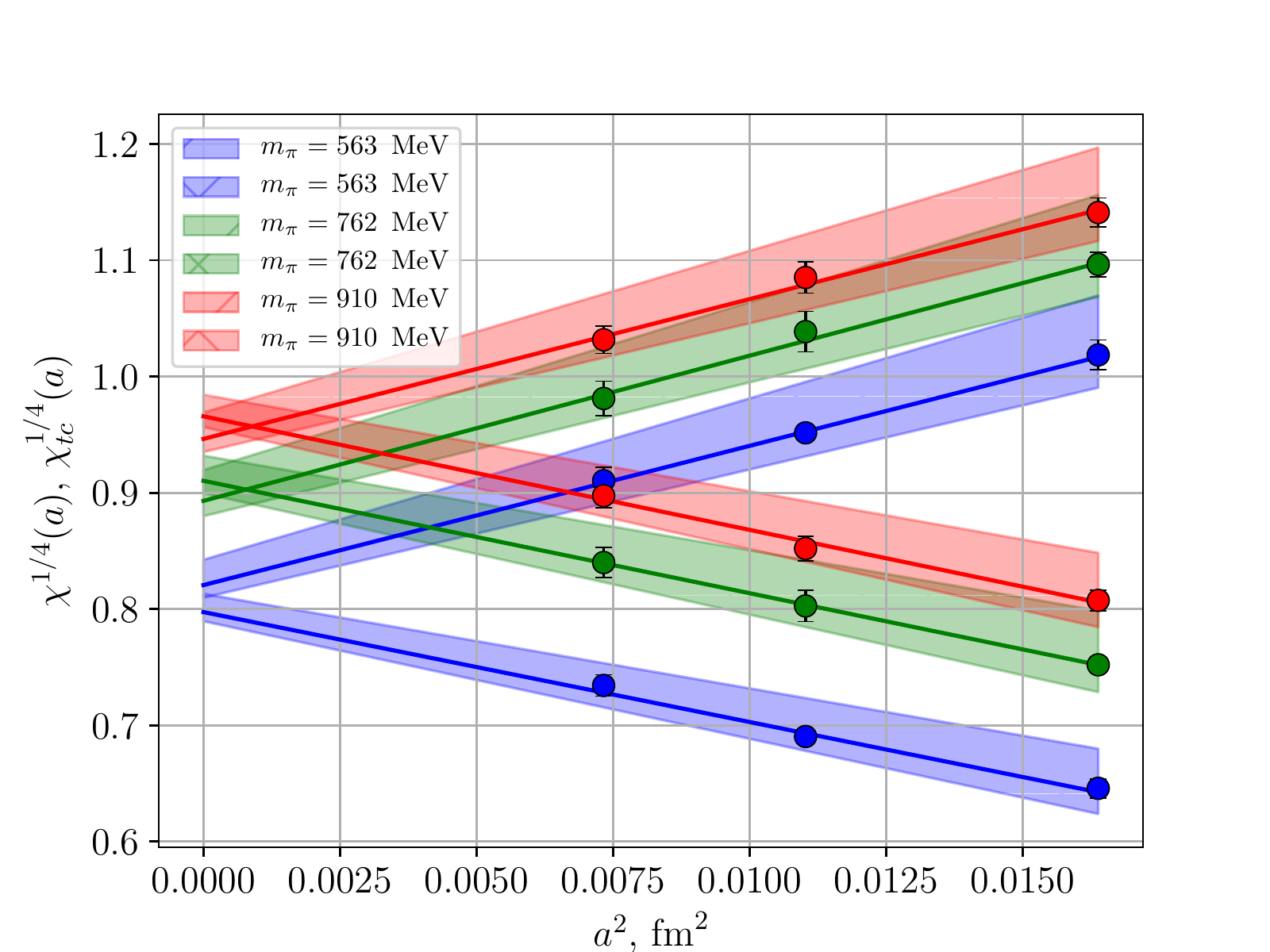}
    \caption{The chiral susceptibility $\chi^{1/4}(a)$ (linear hatching) and the chiral susceptibility weighted in accordance with~\ref{eq:taste_breaking} $\chi^{1/4}_{tc}(a)$ (cross hatching). The continuum extrapolation is done using the fit $\sim A + a^2 B$. Chiral chemical potential is zero $\mu_5=0$.}
    \label{fig:chi_extrapolations}
\end{figure}

In Fig.~\ref{fig:chi_extrapolations} we show the chiral susceptibility $\chi^{1/4}(a)$ (linear hatching) and the chiral susceptibility weighted in accordance with~(\ref{eq:taste_breaking}) $\chi^{1/4}_{tc}(a)$ (cross hatching) for all spacings $a = 0.128\,\mbox{fm}$, $a=0.1054\,\mbox{fm}$, $a=0.0856\,\mbox{fm}$ and all pion masses $563\,\mbox{MeV},\,762\,\mbox{MeV},\,910\,\mbox{MeV}$ for zero chiral chemical potential $\mu_5=0$. In both cases the data are described by the simple $A + a^2 B$ fit with $\chi^2 / \mbox{ndof} < 1$. Note also that the results of the naïve fit of $\chi^{1/4}(a)$ agree with the weighted fit $\chi^{1/4}_{tc}(a)$ in the continuum limit. The same agreement was observed in~\cite{bonati2016axion}. This suggests that the discretisation errors are under control in our study. The same procedure was applied at nonzero chiral chemical potential to get the continuum extrapolated topological susceptibility as a function of $\mu_5$, which is presented in Fig.~\ref{fig:chi}.

Another way to check the discretization errors and also compare result with the ChPT is discussed in~\cite{billeter2004topological}. To do so, we fit the inverse topological susceptibility versus squared pion mass $m_{\pi}^2$ for every fixed lattice spacing $a$ and fixed chical chemical potential $\mu_5$ with the ChPT--motivated anzatz
\begin{equation}\frac{1}{\chi(a)} = \frac{A(a)}{m_{\pi}^2} + B(a).
\label{eq:fit}
\end{equation}

When the fitting parameters $A(a)$ and $B(a)$ are obtained, we perform the continuum extrapolation using the simple square anzatz: 
\begin{equation}\begin{split}
A(a) = A_0 + a^2 A_1,\\
B(a) = B_0 + a^2 B_1.
\end{split}\end{equation}

\begin{figure}[t]
    \centering
    \includegraphics[scale=0.55]{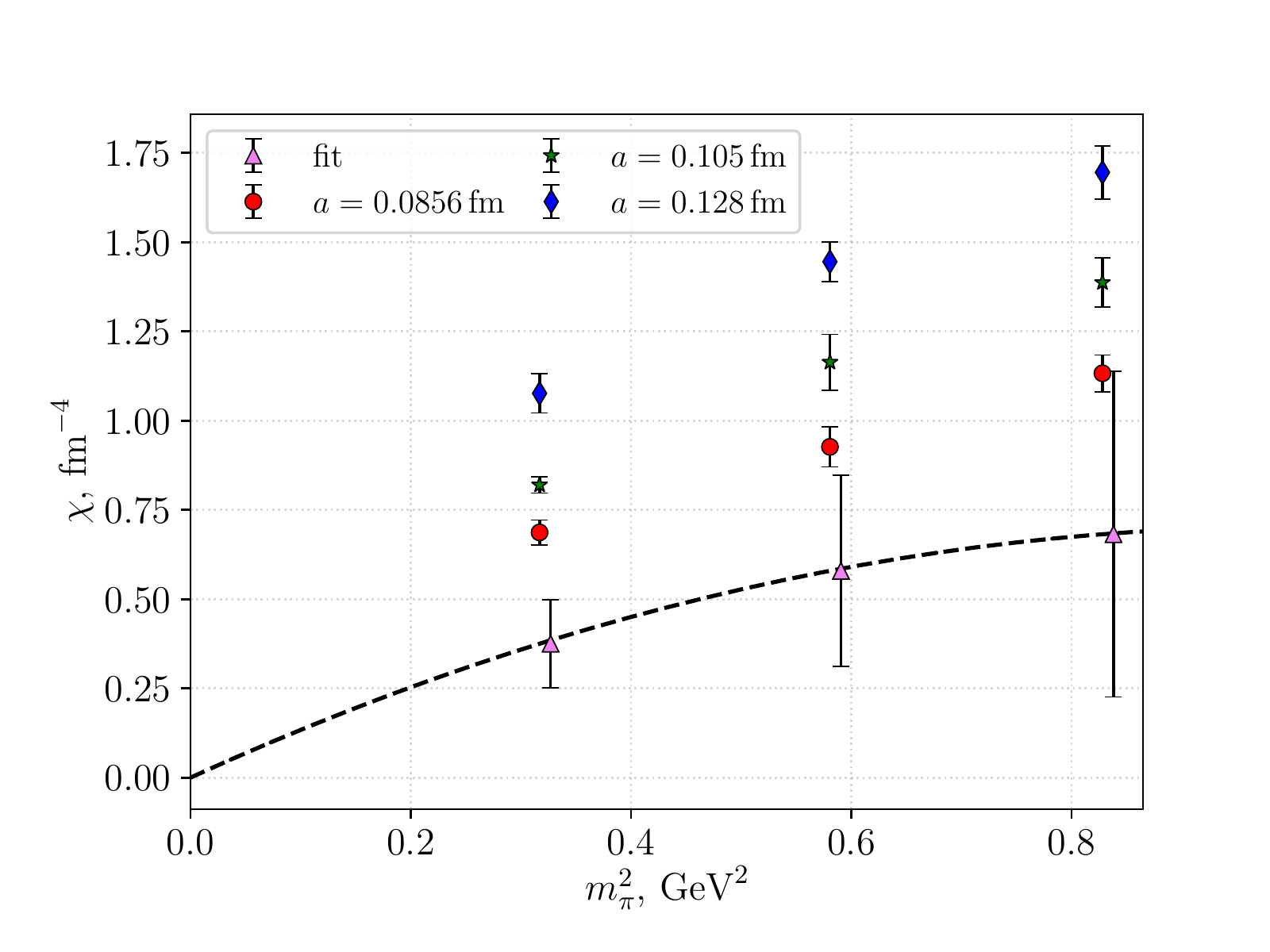}
    \caption{Topological susceptibility at zero chiral chemical potential for all lattice spacings plotted as the function of squared pion mass $m_{\pi}^2$. Violet triangles and dashed line correspond to continuum extrapolation result in accordance with the equation~\ref{eq:fit}.}
    \label{fig:chi_continuum}
\end{figure}

Both fitting stages provide a good description of the data with $\chi^2 / \mbox{ndof} \approx 1$ for all lattice spacings and both $A(a)$ and $B(a)$. We then interpret the function $\chi^{-1}_0 = A_0 / m_{\pi}^2 + B_0$ as the continuum extrapolation of the inverse chiral susceptibility. In Fig.~\ref{fig:chi_continuum} we show our data for all lattice spacings and pion masses together with the continuum extrapolation result. 

The pictures for $\mu_5 > 0$ look the same and thus we do not show them here. From Fig.~\ref{fig:chi_continuum} it is seen that for all pion masses the results in the continuum limit are close to that obtained at the smallest lattice spacing $a = 0.085\,\mbox{fm}$. This allows us to state that the discretization errors are well-controlled within this study. Finally the ChPT states that $A_0=\frac{4}{f_{\pi}^2}$. The extracted value of $f_{\pi}=92(9)$\,MeV, which agrees with the physical $f_{\pi}^{ph}\approx 93$\,MeV.

\end{document}